\documentclass[conference]{IEEEtran}
\usepackage{multirow}
\IEEEoverridecommandlockouts
\usepackage{cite}
\usepackage{amsmath,amssymb,amsfonts}
\usepackage{algorithm}
\usepackage{graphicx}
\usepackage{textcomp}
\usepackage{booktabs} 
\usepackage{tabularx} 
\usepackage{lipsum} 
\usepackage{xcolor}
\usepackage{color,soul}
\usepackage{xspace}
\usepackage{makecell}
\usepackage{balance}
\usepackage{url}

\usepackage[noend]{algpseudocode}

\usepackage{listings}
\def\BibTeX{{\rm B\kern-.05em{\sc i\kern-.025em b}\kern-.08em
    T\kern-.1667em\lower.7ex\hbox{E}\kern-.125emX}}

\newcommand{\agent}{\textit{\textbf{SEEAgent}}\xspace}
\newcommand{\agents}{\textit{\textbf{SEEAgents}}\xspace}

\newcommand{\project}{\textit{eShope}\xspace}

\begin{document}

\title{An LLM-based multi-agent framework for agile effort estimation}



\author{\IEEEauthorblockN{Thanh-Long Bui}
\IEEEauthorblockA{
\textit{Decision System Lab} \\
\textit{School of Computing and IT} \\
\textit{University of Wollongong, Australia}\\
tlb959@uowmail.edu.au}
\and
\IEEEauthorblockN{Hoa Khanh Dam}
\IEEEauthorblockA{
 \textit{Decision System Lab} \\
 \textit{School of Computing and IT} \\
\textit{University of Wollongong, Australia} \\
hoa@uow.edu.au}
\and
\IEEEauthorblockN{Rashina Hoda}
\IEEEauthorblockA{
 \textit{Faculty of Information Technology} \\
\textit{Monash University, Australia} \\
rashina.hoda@monash.edu}
}


\maketitle

\begin{abstract}
Effort estimation is a crucial activity in agile software development, where teams collaboratively review, discuss, and estimate the effort required to complete user stories in a product backlog. Current practices in agile effort estimation heavily rely on subjective assessments, leading to inaccuracies and inconsistencies in the estimates. While recent machine learning-based methods show promising accuracy, they cannot explain or justify their estimates and lack the capability to interact with human team members. Our paper fills this significant gap by leveraging the powerful capabilities of Large Language Models (LLMs). We propose a novel LLM-based multi-agent framework for agile estimation that not only can produce estimates, but also can coordinate, communicate and discuss with human developers and other agents to reach a consensus. Evaluation results on a real-life dataset show that our approach outperforms state-of-the-art techniques across all evaluation metrics in the majority of the cases. Our human study with software development practitioners also demonstrates an overwhelmingly positive experience in collaborating with our agents in agile effort estimation. 



\end{abstract}


\section{Introduction}


Effort estimation is an important aspect of software project management, especially in the planning and monitoring stages. Its accuracy may significantly impact project outcomes -- underestimating effort can lead to missed deadlines and budget overruns, while overestimating can reduce resource efficiency and diminish an organisation's competitive edge \cite{leachSoftwarecostestimation1999}. In modern agile software development, software is built through incremental releases and iterative cycles, each of which requires the completion of a number of \emph{user stories}. As a result, the focus has shifted towards estimating the effort required to complete individual user stories rather than the entire project. It is now standard practice for agile teams to review and estimate each user story in the product backlog, enabling them to plan, prioritise and deliver value efficiently \cite{Pozenel2023AgileEE}.  


At present, agile teams mostly rely on experience, subjective assessments, and consensus to estimate user stories (e.g. planning poker \cite{grenning2002planning}, T-shirt sizing and dot voting). However, these techniques can result in significant inaccuracies, and more notably, inconsistencies across different estimates, even within the same project and the same team \cite{Usman2014}. Recent approaches (e.g. \cite{choetkiertikulDeepLearningModel2019, porruEstimatingStoryPoints2016, scottUsingDevelopersFeatures2018, fuGPT2SPTransformerBasedAgile2023, liFineSEIntegratingSemantic2024}) have proposed a range of artificial intelligence (AI) models that can learn from a team's previous estimates and predict the effort for new user stories. Positive results from those approaches suggest that it is possible to build an AI-enabled prediction system to support agile teams in estimating user stories, enabling them to be consistent in their estimates.  

However, those existing approaches suffer from several major limitations. Most of them are inherent ``black box'' models, and do not provide \textit{explanation} for their estimates. Thus, it is difficult for agile teams to understand them and interpret their estimation. In addition, current agile estimation practices (such as the widely-used Planning Poker \cite{grenning2002planning} method) are \textit{consensus-based}: each team member provides an estimate and a consensus estimate is reached after several rounds of discussions and re-estimation. The purpose here is not only to produce an estimate, but also to promote team conversation and collaboration, leverage the cross-disciplinary nature of agile teams, and help uncover hidden complexities or overlooked aspects of the work. Existing AI-driven estimation approaches lack these important capabilities. They are \emph{not} able to justify their estimates, and \emph{cannot} participate in discussions with human developers to reach an agreed-upon estimate. These critical limitations have resulted in industry reluctance to adopt or deploy them in practice \cite{mohammadChallengesIntegratingArtificial2024}.


This paper aims to address the above limitations by proposing a multi-agent framework for agile estimation. We leverage the recent powerful LLM technologies to build a \emph{new} class of LLM-based agents for agile effort estimation (hereafter called \textbf{S}oftware \textbf{E}ffort \textbf{E}stimation \textbf{Agent} or \agent for short). Each \agent is capable of not only estimating user stories but also providing justification for its estimates. LLMs often suffer from \emph{hallucinations} where the generated information appears plausible but incorrect. To address this problem, we propose a combination of prompt engineering and fine-tuning methods to enrich our agent with knowledge specific to the context of a project through its past user stories. In addition, each agent in our framework can be further enhanced with knowledge specific to its expertise. For example, the front-end agent is enriched with UX knowledge while the back-end agent is with back-end and database code. Thus, our multi-agent approach can combine diverse expertise to lead to more realistic estimates through considering multiple angles of a user story. 

Another key \textit{novelty} of our approach is that our \agent can coordinate and communicate with other agents and human developers in an agile estimation process -- this is a significant departure from the existing approaches. In our multi-agent framework, each individual \agent and human team member can share their perspectives and assumptions on the complexities of a user story before providing their own estimate. If the estimates differ widely, the agents and human team members can discuss the reasons behind the differences. Based on these discussions, our \agent is then able to re-estimate, resulting in a more informed and consensus-based estimate. This may improve the accuracy and reliability of effort estimates by leveraging collective team knowledge and reducing biases. Furthermore, through enabling human-AI collaboration, our approach helps agile teams facilitate conversation, uncover hidden insights, and build team consensus -- a key purpose of agile effort estimation.

We evaluate \agent on two aspects: (i) its accuracy compared to prior state-of-the-art (SOTA) deep learning approaches for estimating user stories; and (ii) how well it works with human developers in the agile estimation process. In terms of accuracy, we used the same dataset and benchmarks as in the SOTA approaches, and the results show our \agent outperforms SOTA in the majority of the cases. With respect to collaborating with humans, we conducted a human study where participants were asked to work with our \agents to estimate user stories for a given software project. Feedback from the participants shows a strong positive experience working effectively with our agents. The participants found that the use of our agents benefit development teams to reach consensus, and bridge the knowledge gap and promote communication between team members. Overall, the participants agreed that they trust our \agent and would recommend using it in practice.

The structure of the paper is as follows. We present the details of \agent in Section~\ref{sec:approach}. The experimental setup and results are reported in Section~\ref{sec:evaluation}. Related work is discussed in Section~\ref{sec:related-work}. Finally, we conclude this paper and outline future work in Section~\ref{sec:conclusion}.




\begin{figure*}[ht]
    \centering
    \includegraphics[width=1\linewidth]{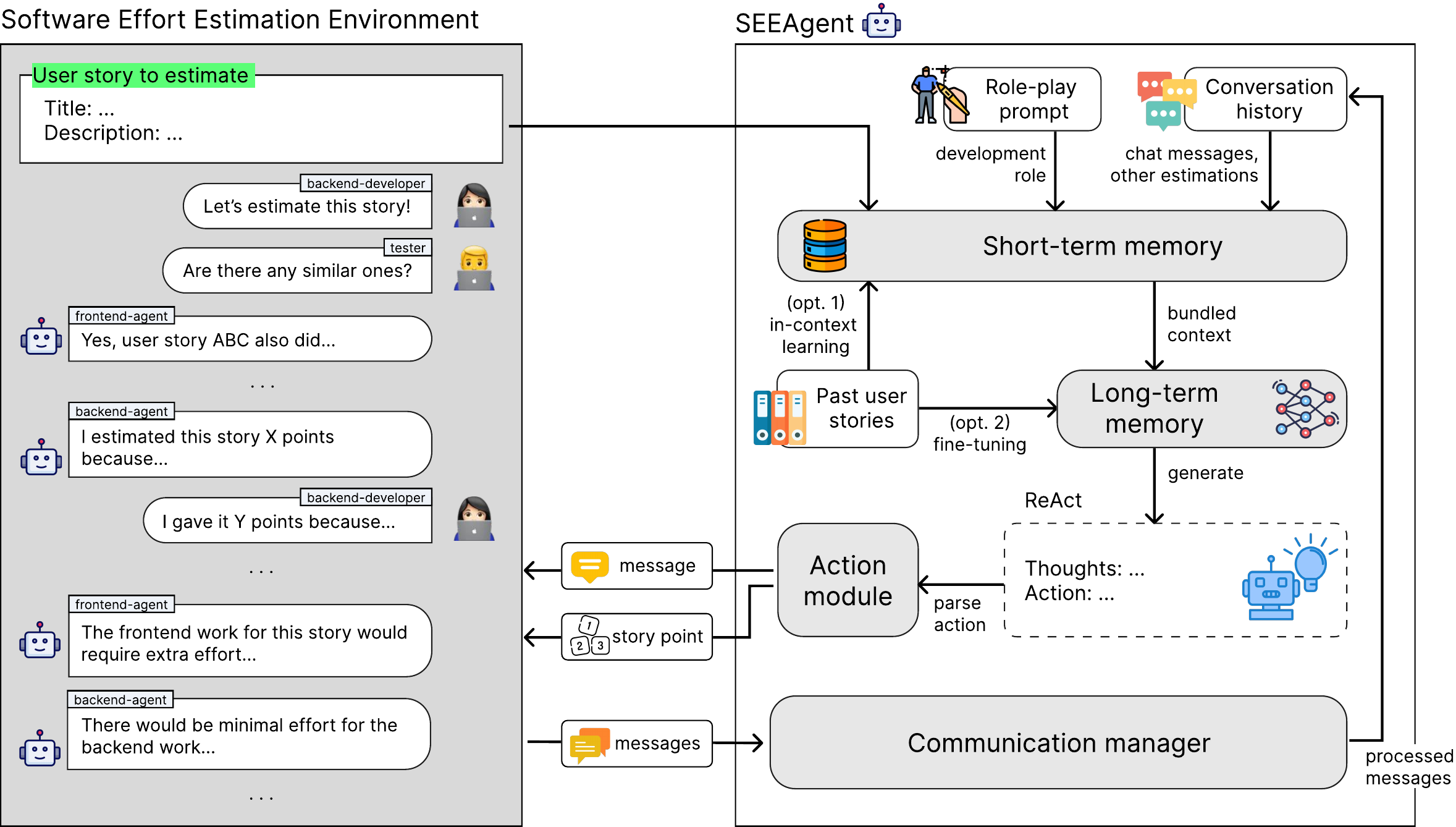}
    \caption{Architecture and example conversations between \agents and humans}
    
    \label{fig:seeagent-framework}
\end{figure*}

\section{Approach}\label{sec:approach}


We propose \textbf{S}oftware \textbf{E}ffort \textbf{E}stimation \textbf{Agent} (\agent), a new agentic framework designed for aiding agile teams in the process of effort estimation and sprint planning  \cite{grenning2002planning}. \agent aims not only to provide accurate and consistent estimates, but also to facilitate informational interactions with agile team members through offering insights and relevant project information beyond mere numerical predictions. 

\subsection{Architectural overview}

We formulate the problem of agile estimation as predicting the effort $E$ (often expressed in terms of \emph{story points}) required to implement a user story described by a natural language specification $Q$. 
Let $\mathcal{D} = \{(Q_i, E_i)\}_{i=1}^{n}$ denote a historical dataset of previously estimated user stories and their associated efforts. Given a new user story $Q'$, the objective is to estimate the required effort $\hat{E}$ such that $\hat{E} \approx E'$, where $E'$ is the unknown true story points.

Each \agent $\mathcal{A}_i$ (see Figure~\ref{fig:seeagent-framework}) operates as an autonomous entity tasked with deriving an effort estimate $\hat{E}$ for a given user story $Q'$. Internally, \agent features a modular architecture comprising four principal components: short-term memory, long-term memory, action module, and communication manager. The agent interacts with the software development environment through multiple means: (i) learns knowledge about projects through fine-tuning \cite{huLoRALowRankAdaptation2021a} or in-context learning on historical user story datasets $D$; (ii) generates estimations for new user stories $Q'$; and (iii) communicates with human practitioners and other agents using its action module. This architecture enables \agent to maintain relevant information, learn from past estimates, and provide informed effort estimates while facilitating interaction with development team members and peer agents.

The internal reasoning process of an agent unfolds over multiple phases, influenced by both local knowledge and peer interactions. Given a user story $Q'$, the agent retrieves relevant information from both memory modules. In short-term memory $\mathcal{M}_s$, the agent stores the user story $Q'$, conversation history $H$, and peer estimates $\hat{E}_j$. From long-term memory $\mathcal{M}_l$, the agent draws relevant examples, patterns, or rationale stored in its latent space.

We follow the Planning Poker practice in agile, where estimating a user story is often conducted in multiple rounds of discussions. At each round $t$, the agent computes its updated effort estimate $\hat{E}^{(t)}$ based on three components:
\begin{itemize}
  \item \textbf{Initial prompt reasoning:} The agent uses $\mathcal{M}_l$ to analyse $Q'$ and determines whether clarification questions are needed before making an estimate.
  
  \item \textbf{Contextual feedback integration:} For $t > 0$, the agent observes peer predictions $\{\hat{E}_j^{(t-1)}\}_{j \ne i}$ and prior conversation trace $H^{(t-1)}$, then uses $\mathcal{M}_l$ to analyse them and determines if additional clarification questions are needed before updating its estimate.
  
  \item \textbf{Effort generation:} Based on available information and clarifications, the estimate is derived:
  $$
  \hat{E}_i^{(t)} = \mathcal{M}_l(\cdot|Q', \mathcal{M}_s, \{\hat{E}_j^{(t-1)}\}_{j \ne i}, H^{(t-1)})
  $$
\end{itemize}

The agent continues this process until convergence is detected across participants or the maximum round limit $r$ is reached. 



This architecture allows agent behaviour to be explainable and auditable, with the communication trace and reasoning process offering insight into how $\hat{E}_i^{(t)}$ was derived. We will now discuss the major components of our \agent architecture in detail. 

\subsection{Long-term memory}

The long-term memory component, with reasoning capabilities powered by general-purpose LLMs, serves as the central information processing unit. It pertains to project-specific knowledge and general world knowledge. Given the stateless nature of language models \cite{sumersCognitiveArchitecturesLanguage2024}, the information in long-term memory is prone to being outdated, necessitating continual updates. To address this, the process of enriching the module with the latest information about the project must be routinely updated through fine-tuning using newly estimated user stories. This continuous refinement ensures that it remains aligned with evolving project requirements. Furthermore, the integration of fresh information helps the agent to be aware of changes and provide more informed and context-aware responses, thereby making them more useful in supporting human practitioners in agile estimation sessions.

Story point estimation is inherently project-relative \cite{choetkiertikulDeepLearningModel2019}, with identical user stories potentially receiving different effort estimations across development teams. Therefore, to accommodate this variability, our approach adapts the base LLM model through two modes: fine-tuning for established projects with sufficient historical data, and in-context learning \cite{brownfewshotlearners2020} for new projects. The fine-tuning process is detailed below, while the in-context learning process is discussed in Section \ref{sect:short-term-memory}.


Large model fine-tuning is computationally costly and requires a substantial amount of data and computational resources. Due to the constraint of computation resources, we leveraged parameter-efficient fine-tuning (PEFT) techniques (e.g. \cite{pfeifferMADXAdapterBasedFramework2020, huLoRALowRankAdaptation2021a, houlsbyParameterEfficientTransferLearning2019, heUnifiedViewParameterEfficient2022, liPrefixTuningOptimizingContinuous2021}). Specifically, we employed Low-rank Adaptation (LoRA) \cite{huLoRALowRankAdaptation2021a} to adjust the chosen base model's weights. LoRA introduces new trainable parameters in the form of low-rank decomposition matrices to the attention layers and freezes the pre-trained model weights. This allows the model to learn project-specific information while drastically reducing the number of trainable parameters. To further reduce the computation requirements, we applied QLoRA \cite{qlora2023}, 4-bit quantisation in combination with LoRA. QLoRA compresses the model weight precision from BF16 to NF4, helping reduce the memory usage four times, in theory, without significantly impacting the performance of the model. 

In this study, we employed the \textit{Llama-3.1-8B-Instruct} model as our base model. After LoRA is applied to the model, 20,971,520 additional parameters are introduced while the original 8 billion parameters are frozen. These newly introduced parameters are then updated through the Supervised Fine-tuning (SFT) technique \cite{ouyangTrainingLanguageModels2022a} trained on historical user stories with ground-truth estimates. 
SFT allows the model to learn downstream tasks by training it on labelled input-output pairs relevant to the tasks. The input will be a structured prompt containing a user story, and its story point will be the output.

The effectiveness of fine-tuning relies on the quality and design of the prompt used during training. Since our objective is to teach the base model to estimate story points for agile effort estimation, we crafted a prompt that closely resembles how a human practitioner would respond to such queries. The prompt is structured following a conversation format with clearly separated system instructions, user input (the user story), and assistant answer (the estimated story points). This structure enables the model to learn the semantic relationship between a user story and its corresponding numerical estimate while preserving the conversational context essential for agent and human interactions.

The system message sets the context by assigning a specific role to the model: an expert software development effort estimator. The user message presents the user story, and the assistant message entails a succinct response, making it easier for the model to learn from examples. Through training with this prompt structure across multiple examples with varying story complexities and associated effort points, the model gradually learns to associate different types of user stories with appropriate effort levels. The complete prompt format used for fine-tuning is shown in Listing~\ref{lst:finetune-prompt}.
\vspace{0.5cm}


\begin{lstlisting}[language=, caption={Prompt for fine-tuning model}, label=lst:finetune-prompt]
[system]
You are an expert software development effort estimator. Given a software development issue summary, predict the effort in story points.

[user]
### Summary:
{{USER STORY}}

[assistant]
My estimated story point is: {{ESTIMATED POINT}}
\end{lstlisting}

\subsection{Short-term memory}\label{sect:short-term-memory}

Short-term memory functions as immediate information storage for an estimation session. It maintains a context consisting of a development role, recently received messages, reasoning states, and relevant project artifacts. All these data are structured following the ReAct \cite{yao2023react} prompting template. This template leverages interleaving explicit thoughts, proposed actions, and observations, making the agent's thought process auditable and traceable. Furthermore, using this prompting technique allows the agent to carry out autonomous decision-making and action-taking capabilities imposed by the implemented language model. 


\begin{lstlisting}[numbers=left,numbersep=10pt,xleftmargin=15pt,basicstyle=\ttfamily\footnotesize,numberstyle=\tiny\color{gray},caption={SEEAgent system prompt}, label=lst:system-prompt]
{{ROLE-PLAY PROMPT}}

You are an agent that participates in a planning poker session with other agents and human players. 

## Actions
You have access to actions that you can use. Think carefully about which actions to use.

Available actions:
{{ACTIONS DESCRIPTION}}

## Input format
You will receive game state and chat history updates.

## Output format
Always follow this format:

Thought: your reasoning about what to do next
Action: the_action_name
Action Input: {{"param1": "value1", "param2": "value2"}}

## Planning poker rules
{{GAME RULE}}

## Strategy
You objective is to communicate via chat with other players to come up with the best estimation for the given user story. For better estimation, you should base your estimation on the past user stories and similar user stories.

In the first round, you can either make an estimation or ask questions to clarify the user story. For the following rounds, you should justify your estimation based on similar user stories and consider others' estimations.

## Past user stories
For reference, you can use the following past user stories to help you make your estimation:

{{PAST USER STORIES}}

## User story to be estimated
{{USER STORY}}

\end{lstlisting}

The system prompt of \agent in Listing~\ref{lst:system-prompt} comprises several sections:

\begin{itemize}
    \item Line 1-3: The role-play prompt assigns specific profiles in software development to an agent, influencing the agent's responses to match the desired expertise. This role-play configuration consists of two principal elements: a dynamic software development expertise profile configured during agent initialisation, and a static profile defining \agent's fundamental purpose as a planning poker session participant.
    \item Line 5-19: The prompt uses the ReAct prompting technique \cite{yao2023react} through dedicated sections for actions, input specifications, and output formatting.
    \item Line 21-27: To ensure appropriate agent behaviour within the designed planning poker environment, dedicated sections detail the operational rules and strategic guidelines. These sections provide the necessary behavioural constraints and decision-making frameworks.
    \item Line 29-32: Our approach includes an optional in-context learning component through the incorporation of past user stories, providing an alternative approach when LLM fine-tuning is not feasible or when the project is in its early stage and the number of past user stories with ground-truths is limited.
    \item Line 34-35: The final section details the target user story designated for effort estimation.
\end{itemize}


\subsection{Action and communication}

The grounding of an \agent is conducted in part by its Action module, which enables it to interact with the software development environment. The Action module acts as an intermediary between the internal of the agent, the reasoning engine, and the environment. This module is integral to \agent's ability to elicit information and respond to stakeholders, ensuring efficient communication with human practitioners and other agents. The action module defines two primitive interaction actions: (a) $chat$, allowing the agent to provide justifications, engage in collaborative dialogue, and articulate reasoning processes; (b) $make\_estimation$, allowing the agent to commit to an estimation. The input interfaces of these actions are concisely predefined to ensure the tool-use capability of LLMs can efficiently use them.

The Communication Manager serves as the main interface for managing external communication. The centralised message aggregation method implemented by this module handles communications from the environment, e.g. human practitioners and other \agents, and pre-processes them before passing them into the short-term memory. This module ensures operational coherence by systematically coordinating messages between external sources and the internal mechanisms of an agent. The role of this module is necessary as it provides a structured and meaningful stream of information that allows the agent to communicate effectively.

\begin{figure}
    \centering
    \includegraphics[width=1\linewidth]{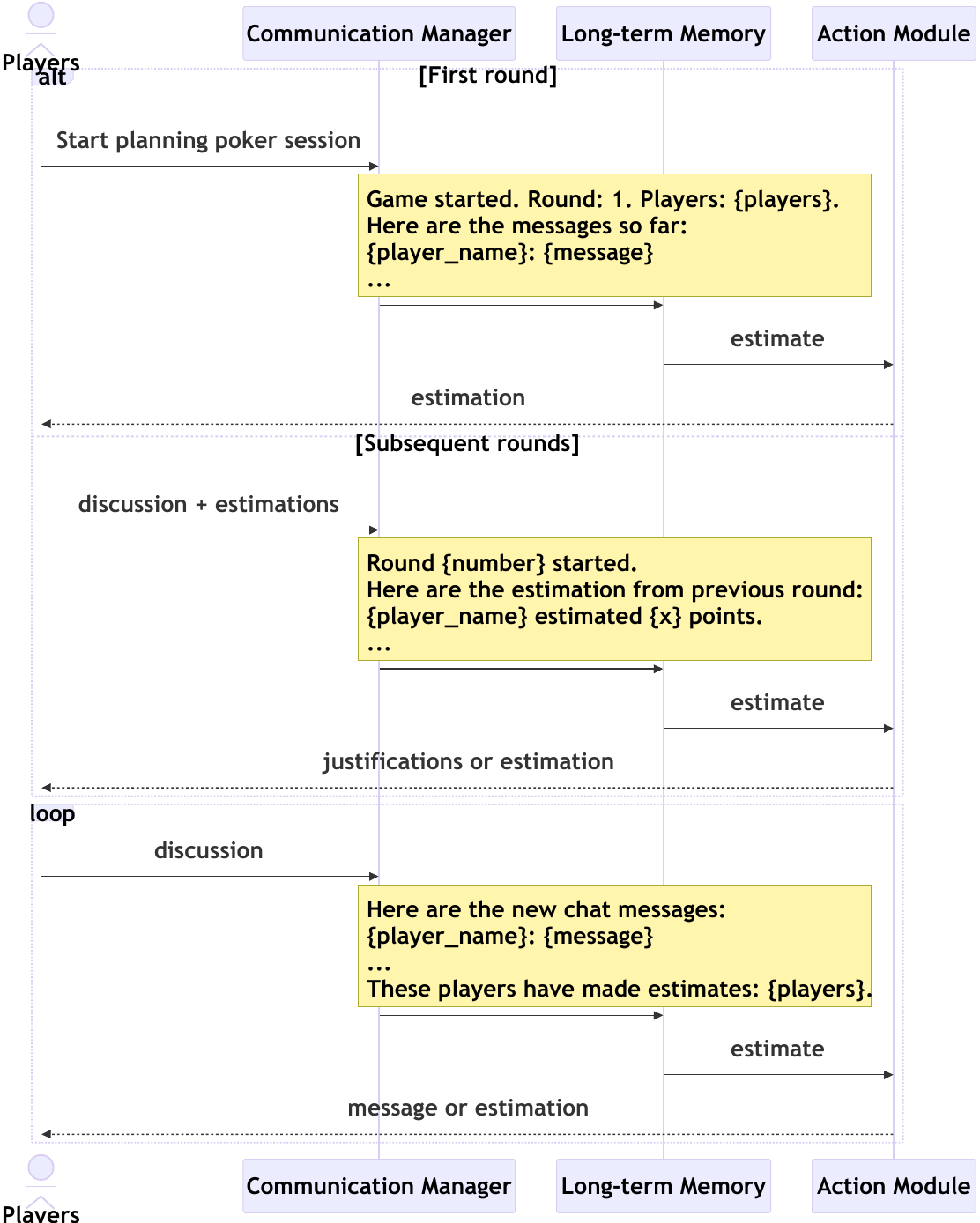}
    \caption{Communication protocol of \agent in an agile estimation session based on the Agile Planning Poker style}
    \label{fig:message-protocol}
\end{figure}

The communication protocol of \agent, illustrated in Figure~\ref{fig:message-protocol}, follows a structured communication flow between multiple internal components. The protocol differentiates between the first round and the subsequent rounds of interaction. On initiation of the first round, the Communication Manager aggregates players' information and message content into a structured prompt, whereas in subsequent rounds, the Communication Manager incorporates prior round estimations. This structured prompt is passed into long-term memory, enabling the agent to reason and determine its next action -- whether to make an estimate or facilitate discussion. The agent then enters a loop of discussion with the players, and estimates are refined until the round concludes.

\subsection{Prototype implementation}


We have implemented our framework in a platform for human developers and \agents to work together in estimating user stories for a given software project. The platform consists of a back-end server, which is responsible for orchestrating the agile Planning Poker estimation session mechanics and real-time updates. This component relies on the WebSocket protocol \cite{melnikovWebSocketProtocol2011} to establish bidirectional communication with the Agent server and front-end. The protocol enables real-time broadcasting of participant actions across all connected clients. The Agent server functions as the centralised management system for \agents. This component is responsible for agent initiation and establishing WebSocket-based communication with the back-end server. Upon initialisation, each agent is configured with a specific role-play prompt. The system allows for flexible model selection to be used as the long-term memory, with the \textit{gpt-4o-mini} as the default one.

\begin{figure}[ht]
    \centering
    \fbox{
    \includegraphics[width=1\linewidth]{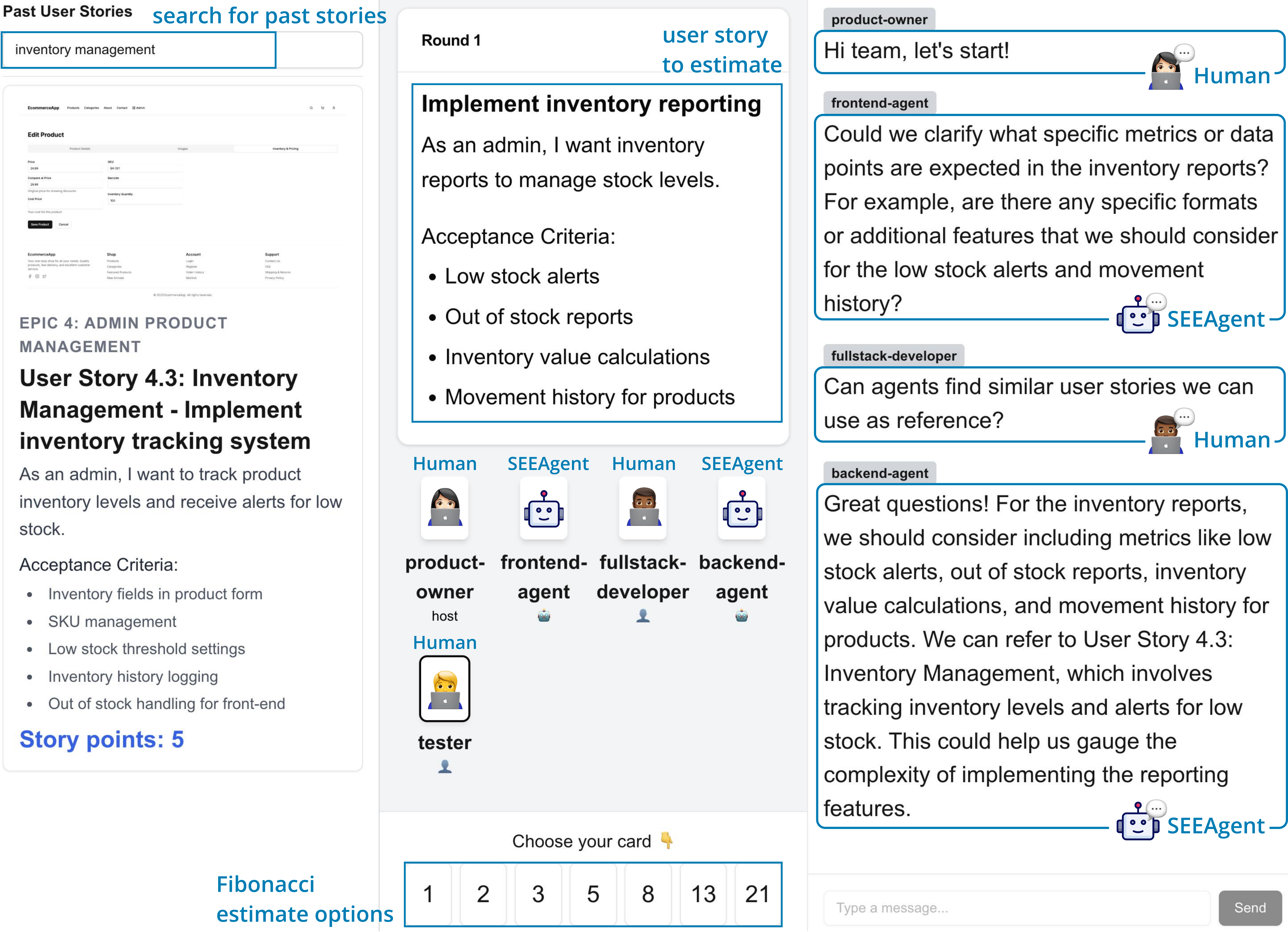}}
    \caption{Screenshot of the front-end for human players}
    \label{fig:frontend}
\end{figure}


The front-end is implemented as a web-based graphical user interface that facilitates human-AI interaction in a typical agile estimation based on the widely used Planning Poker method. Figure~\ref{fig:frontend} illustrates the interface presented to the users. The middle column contains a card showing the active estimation round and the current user story being estimated. A roster where the active estimator is visually distinguished through a black border indication. Positioned at the bottom of the central panel, there is a Fibonacci sequence for point allocation. In the left panel, users are presented with past user stories of a software project developed for the human study. A text-based search box is included for quickly finding the stories that the agents refer to. The right panel incorporates a real-time chat interface, facilitating message exchange between human participants and agents.

More details of our \agent platform can be found at: \textit{\url{https://figshare.com/s/bb33c350859802b071ea}}.

\section{Evaluation}\label{sec:evaluation}

Our evaluation aims to demonstrate the predictive performance of a single \agent against state-of-the-art techniques for agile effort estimation. We also examine how these agents synergise within human teams in a designed environment through a comprehensive human study. This dual approach allows us to assess both estimation accuracy and practical adoption factors of our framework. The research questions we aim to answer are: 


\begin{itemize}    
    \item \textbf{RQ1:} \textit{How accurate is \agent in estimating user stories compared to state-of-the-art (SOTA) techniques?}

    Our agent architecture relies on long-term memory as the source to produce the estimations; hence, it is critical to validate whether it can churn out accurate effort estimates. Accurate estimates are vital for the agent's effectiveness in planning sessions as they build credibility in team discussions and enable the agent to meaningfully contribute to the conversation. To evaluate this integral component, we compare our approach against the most recent deep learning approaches, namely: Deep-SE \cite{choetkiertikulDeepLearningModel2019}, GPT2SP \cite{fuGPT2SPTransformerBasedAgile2023}, and Fine-SE \cite{liFineSEIntegratingSemantic2024}. The same evaluation metrics (mean absolute error, mean magnitude of relative error, and performance indicator) and datasets in the SOTA approaches are used for a fair comparison. This question is essential to ensure our architectural choice of using LLMs for estimation is valid.

    
    \item \textbf{RQ2:} \textit{How do human agile developers perceive  \agent in working with our \agents in agile effort estimation?}

    For AI agents to be effective collaborators, they must be accepted and trusted by development teams. Therefore, understanding how human practitioners perceive and interact with our agents within an effort estimation environment is vital -- it reveals the benefits, challenges, and improvement areas where the agents might possess. To answer this question, we conduct a human study with software development professionals, letting them interact with agents in a designed environment and then collect their feedback through a mixed-format survey. This human-centred evaluation provides us with insights into the practical viability of our approach and guide improvements for real-world adoption.


\end{itemize}

\subsection{\textbf{RQ1: Effectiveness of \agent in estimating user stories}}

\subsubsection{Dataset} 

We used a publicly available dataset from the Fine-SE paper \cite{liFineSEIntegratingSemantic2024}. In the paper, the authors trained and evaluated their approach on a dataset with 17 industrial projects and four open-source software projects. To allow direct comparison with their approach and ensure reproducibility, we utilised their publicly available dataset of four open-source projects: Mesos (ME), Data Management (DM), Talend Data Quality (TD), and Usergrid (US). The data was mined from the Jira issue tracking system of where these projects are managed. The statistical descriptions of the project are presented in Table~\ref{table:opensource-dataset}. 

\begin{table}[h]
\centering
\caption{Statistical description of the open-source projects}
\label{table:opensource-dataset}
\begin{tabular}{|c|r|ccccc|}
\hline
Project & No. stories & \multicolumn{5}{c|}{Story points} \\
\cline{3-7}
& & Min & Max & Mean & Median & Std \\
\hline
DM & 4,320 & 0 & 195 & 7.91 & 3 & 16 \\
ME & 414 & 1 & 13 & 3.93 & 3 & 2.6 \\
TD & 73 & 0.5 & 40 & 8.25 & 8 & 7.4 \\
US & 100 & 1 & 8 & 3.31 & 3 & 1.2 \\
\hline
Total & 4,907 & - & - & - & - & - \\
\hline
\end{tabular}
\end{table}

\subsubsection{Baselines} 

In part of the study, we experiment with two variants of \agents: one utilised the base \textit{Llama-3.1-8B-instruct} model, and another implemented a fine-tuned version of the same model for the SEE task. LLMs have demonstrated significant abilities to understand the semantics in natural language, making them potential candidates for tasks involving the understanding of textual information in software tasks. While the base model \agent leverages the general knowledge encoded in its parameters, the fine-tuned variant combines this foundational knowledge with project-specific insights gained through additional training on SEE-specific data.


In addition, we compare the performance of our approach against state-of-the-art techniques in agile effort estimation, including Fine-SE \cite{liFineSEIntegratingSemantic2024}, GPT2SP \cite{fuGPT2SPTransformerBasedAgile2023}, and Deep-SE \cite{choetkiertikulDeepLearningModel2019}. Deep-SE established the initial adoption of deep learning for SEE when it introduced a hybrid architecture combining Long Short-term Memory (LSTM) networks with Recurrent Highway Networks (RHN). GPT2SP extended the GPT-2 architecture to generate point estimates. Most recently, Fine-SE employed BERT \cite{devlinBERTPretrainingDeep2019} to extract semantics features from user stories, integrating these with neural networks to produce effort estimates. We have replicated the evaluation of those approaches based on the code they provided. 





\subsubsection{Implementation configurations}

In this experiment, we focus solely on evaluating the long-term memory component of \agent, comparing the accuracy of two variants: one using the base \textit{Llama-3.1-8B-instruct} model, and another using its fine-tuned version for the SEE task. We evaluate \agent in this standalone configuration to ensure a fair comparison, since \agent's full architecture offers additional capabilities through multi-agent communication and interactive refinement. We deliberately isolate the core estimation component to maintain equivalent testing conditions with previous approaches. This controlled setup enables direct performance comparison on the same metrics and datasets used.  


For each project in our dataset, we performed separate fine-tuning of the base LLM model. We utilised the Unsloth library \cite{unsloth} for this process as it supports the QLoRA fine-tuning technique. To ensure fair comparison with results from the Fine-SE paper, we maintained the same train-test split ratios and data distribution as used in their published implementation. The fine-tuning process was consistent across all projects: we configured the loading of the model into memory with the following parameters: $\textit{max\_sequence\_length}$ is set to 2048 tokens; the LoRA rank was set to 8 with an equal scaling factor to maintain a balance between the model capacity and new knowledge introduction. For training, we configured the epoch to 10 with a learning rate of \textit{2e-4}. The training processes were conducted using Google Colab's A100 GPU infrastructure.

\subsubsection{Evaluation metrics}

We use the same evaluation metrics, i.e. mean absolute error (MAE), mean magnitude of relative error (MMRE), and performance indicator (PRED), as in a prior work \cite{liFineSEIntegratingSemantic2024}. MAE measures the average of absolute differences between predicted and ground truth estimation. It intuitively captures the proposed model's average error margin across the whole test set. 

\begin{equation}
\label{mae_equation}
\text{MAE} = \frac{1}{n} \sum_{i=1}^{n} |y_i - \hat{y}_i|
\end{equation}

where $y_i$ and $\hat{y}_i$ are the ground truth estimation and the predicted estimation, respectively. $n$ is the number of instances in the test set.

MMRE computes the average magnitude of the relative error between the predicted and ground-truth estimations. It presents the relative normalised magnitude of error made by the model on the given projects, which is ideal for comparison across multiple projects of different estimation units systems.

\begin{equation}
\label{mmre_equation}
\text{MMRE} = \frac{1}{n} \sum_{i=1}^{n} \frac{|y_i - \hat{y}_i|}{y_i}
\end{equation}

PRED(50) examines the percentage of predictions that fall within 50\% of ground truth estimation values. It offers a practical insight by showing how often the model achieves a certain level of accuracy, highlighting the model's ability to make estimations within a threshold of acceptable error margin (50\%).

\begin{equation}
\label{pred50_equation}
\text{PRED}(50) = \frac{1}{n} \sum_{i=1}^{n} \begin{cases} 1, & \text{if } \frac{|y_i - \hat{y}_i|}{y_i} \leq 0.50 \\ 0, & \text{otherwise} \end{cases}
\end{equation}

\subsubsection{Results}

\begin{table}[!h]
\caption{Benchmark results of \agent vs baselines}
\label{table:finetune_results}
\centering
\begin{tabular}{llrrr}
\hline
Proj & Approach & MAE & MMRE & PRED(50) \\
\hline
DM & Deep-SE & 4.786 & 3.118 & 0.315 \\
   & GPT2SP & 4.303 & 1.900 & 0.362 \\
   & Fine-SE & 4.628 & 1.739 & 0.264 \\
   & $\agent_\text{base model}$ & 17.053 & 11.698 & 0.157 \\
   & $\agent_\text{fine-tuned}$ & \textbf{4.251} & \textbf{1.491} & \textbf{0.383} \\
\hline
ME & Deep-SE & 1.382 & 0.489 & 0.687 \\
   & GPT2SP & 1.583 & 0.629 & 0.578 \\
   & Fine-SE & 1.395 & 0.449 & 0.723 \\
   & $\agent_\text{base model}$ & 7.398 & 2.463 & 0.108 \\
   & $\agent_\text{fine-tuned}$ & \textbf{1.217} & \textbf{0.414} & \textbf{0.747} \\
\hline
TD & Deep-SE & \textbf{4.215} & 1.121 & \textbf{0.533} \\
   & GPT2SP & 5.736 & 2.231 & 0.400 \\
   & Fine-SE & 4.659 & \textbf{0.755} & 0.400 \\
   & $\agent_\text{base model}$ & 7.133 & 1.382 & 0.467 \\
   & $\agent_\text{fine-tuned}$ & 5.133 & 1.456 & 0.333 \\
\hline
US & Deep-SE & 0.618 & 0.163 & 0.850 \\
   & GPT2SP & 0.447 & 0.120 & 0.950 \\
   & Fine-SE & 1.826 & 0.541 & 0.150 \\
   & $\agent_\text{base model}$ & 11.300 & 2.797 & 0.050 \\
   & $\agent_\text{fine-tuned}$ & \textbf{0.350} & \textbf{0.085} & \textbf{1.000} \\
\hline
\end{tabular}
\begin{minipage}{\textwidth}
\vspace{2mm}
\footnotesize
\begin{tabbing}
\hspace{10pt} \= \textsuperscript{1} The lower of MSE and MMRE, the better.\\
\> \textsuperscript{2} For PRED(50), a higher value is better.\\
\> \textsuperscript{3} The best results on evaluation measures are highlighted in bold.
\end{tabbing}
\end{minipage}
\end{table}

The benchmark results (see Table~\ref{table:finetune_results}) between $\agent_\text{base model}$ and $\agent_\text{fine-tuned}$ reveal remarkable performance differences across all projects. For the DM project, the base model shows considerable deviation with an MAE of 17.053 and MMRE of 11.698, whereas the fine-tuned version substantially reduces these errors to 4.251 and 1.491, respectively. In the ME project, the fine-tuned variant improved MAE by 83.55\% (from 7.398 to 1.217) and MMRE by 83.18\% (from 2.463 to 0.414). The most improvement is observed in the US project, where fine-tuning reduces MAE from 11.3 to 0.35 and MMRE from 2.797 to 0.085. In terms of the PRED(50) metric, the fine-tuned model achieves notably higher scores across all projects, with 591\% improvement in ME (from 0.108 to 0.747) and a 20-fold increase for US (from 0.05 to 1). These consistent improvements across the projects strongly suggest that the fine-tuning process successfully enabled the model to learn project-specific effort estimation patterns. 

When comparing against the current state-of-the-art (SOTA) approaches, $\agent_\text{fine-tuned}$ outperforms in three out of four projects.
In the DM project, our fine-tuned agent achieves up to 16.63\% improvement over the second-best results across the evaluation metrics, notably reducing the MMRE from 1.739 down to 1.491. 
Regarding the ME project, our approach yields superior results across all metrics (MAE: 1.217, MMRE: 0.414, PRED(50): 0.747), outperforming Deep-SE (1.382, 0.489, 0.687), GPT2SP (1.583, 0.629, 0.578), and Fine-SE (1.395, 0.449, 0.723). 
For the US project, our fine-tuned agent achieves notable performance with the lowest MAE (0.350) and MMRE (0.085) among all approaches while having the perfect PRED(50) score. 
As for the TD project, our approach (MAE: 5.133, MMRE: 1.456, PRED(50): 0.333) outperforms GPT2SP across all metrics, but still leaves room for improvement compared to Deep-SE (MAE: 4.108, PRED(50): 0.438) and Fine-SE (MMRE: 0.692). 

\begin{table}[h]
\centering
\caption{Comparison Between Approaches Using Wilcoxon Test and $\hat{A}_{12}$ Effect Size (in Brackets)}
\begin{tabular}{l|rrr}
\hline
\agent vs & Deep-SE & GPT2SP & Fine-SE \\
\hline
DM & $<$0.001 [0.59] & 0.061 [0.53] & 0.007 [0.55] \\
ME & 0.028 [0.65] & 0.042 [0.63] & 0.032 [0.67] \\
TD & 0.095 [0.27] & 0.359 [0.73] & 0.524 [0.40] \\
US & $<$0.001 [0.95] & 0.002 [0.90] & $<$0.001 [0.95] \\
\hline
\end{tabular}
\label{tab:wilcoxon}
\end{table}


Table~\ref{tab:wilcoxon} presents the results of the Wilcoxon test (along with Vargha-Delaney $\hat{A}_{12}$ effect size) to measure the statistical significance of $\agent_\text{fine-tuned}$ against the SOTA approaches. Using the alpha level $\alpha$ of 0,05, our approach displays the strongest results in the US project, demonstrating statistical significance (p $<$ 0.001) and a large effect size (0.9 to 0.95) compared to all other approaches. Performance in DM and ME projects shows moderate improvements with small to medium effect sizes (0.53 to 0.67), while results in the TD project indicate no significant improvements.




\subsection{\textbf{RQ2: Human-AI collaboration}}

\subsubsection{Experimental design and procedures}




We conducted a human user study using a mixed-format survey approach that combines interactive prototype evaluation with subsequent data collection through a survey incorporating both structured (quantitative) and open-ended (qualitative) response formats. The study was approved by our university's Human Research Ethics Committee (approval number 2025/015).

There are 12 participants recruited through professional networks within academic and industry settings, utilising a purposive sampling approach to ensure participants possessed relevant software development experience. To maintain ethical standards and minimise potential bias, all prospective participants were explicitly informed that their participation was voluntary and their decision to participate or decline would not affect their professional or personal relationships with the researchers.


The participants are asked to join a session in which they work collaboratively with some other human participants and \agents to estimate user stories for a given software project. We chose a small, simplified e-commerce software project (namely \project) which has functionalities (e.g. log in, log out, shopping cart, order, etc.) that are familiar to all the participants.  The \project project consists of 33 user stories written based on the Connextra format \cite{10.1007/978-3-319-30282-9_14} and containing several acceptance criteria. Details of this project can be found at \textit{\url{https://figshare.com/s/592bb84543653d80a515}}. 

Each estimation session involves six participants: three human developers, two \agents with different development roles, and a human product owner (the first author) who presents the user stories and guides the session. During each session, participants will estimate various user stories of the \project application. The estimation session is designed to last approximately 30 minutes. Participants are then asked to complete a survey that consists of 12 Likert-scale questions and two open-ended questions.

The estimation session is designed based on the agile Planning Poker method (also known as Scrum Poker), a collaborative estimation technique widely used by agile teams in the industry \cite{grenning2002planning}. The process begins with the product owner introducing a user story that requires estimation. The participants (\emph{including both \agents and human developers}) then discuss the details of the story, ensuring everyone understands the requirements and scope. Each participant will then provide their estimate in story points. Next, the estimates are compared. If all estimates are the same, indicating consensus, the team moves to the next story. However, if there is a difference in estimates, the participants provide justifications for their estimates, and a group discussion follows to address any misunderstandings or differing perspectives. After this discussion, the participants re-estimate the story. This cycle of discussion and re-estimation continues until the team reaches a consensus. The process ensures that the team reaches a shared understanding of the story's requirements and agrees on the estimated effort.

\subsubsection{Implementation configurations}

Since the number of user stories of \project is limited, instead of fine-tuning the long-term memory of \agent, we opted for the option of using in-context learning \cite{brownfewshotlearners2020}. This method incorporates a small number of ground-truth samples directly into the short-term memory. This approach enables us to leverage the model's existing capabilities while contextually adapting to the project-specific knowledge without extensive training data. We used the \textit{gpt-4o-mini} model from OpenAI as the long-term memory for \agent as it provides a good balance between response time and reasoning ability. 


We implemented two variants of \agent with role-play prompting following research demonstrating that LLMs can effectively embody specialised technical roles. \cite{kongBetterZeroShotReasoning2024}. Each variant was initialised with either front-end development or back-end development expertise. This is achieved by establishing a role in agents' short-term memory when initiating them: \textit{You are a \{role\} developer agent}. This methodology is backed by empirical evidence \cite{kongBetterZeroShotReasoning2024} suggesting that role-play prompting helps models maintain consistent domain expertise and generates more effective domain-specific reasoning compared to without the prompt. The role-specific agents exhibited different technical perspectives aligned with their assigned development domains, enabling the study participants to gather insights from complementary areas of software engineering expertise.


\subsubsection{Post-session survey}

We designed a mixed-format survey (see Table~\ref{table:survey})  to gain a holistic view of human practitioners' perception and acceptance toward \agents. The survey has a structured questionnaire that assesses multiple aspects of \agent, including its ability to facilitate consensus, bridge knowledge gaps, identify risks, ease of use, and quality of collaboration with human developers. The survey also evaluated practitioners' sentiment towards the usage of \agents in SEE, including their trust in the agents' estimations, willingness to rely on and recommend such systems. This survey helps evaluate both the practical utility and adoption potential of \agents in SEE contexts. The quantitative analysis was complemented by qualitative data from open questions, which helps explore perceived benefits, challenges, and potential improvements to better agent design in software development environments.

\begin{table}[ht]
\caption{Human Study Questionnaire}
\label{table:survey}
\centering
\renewcommand{\arraystretch}{1}
\begin{tabular}{|l|p{7.5cm}|}
\hline
\textbf{ID} & \textbf{Question} \\
\hline
\multicolumn{2}{|l|}{\textbf{Perceptions of \agents}} \\
\hline
Q1\textsuperscript{\ddag} & The AI agent’s estimations helped us reach consensus efficiently \\
Q2\textsuperscript{\ddag} & I think AI agents can bridge knowledge gaps between team members \\
Q3\textsuperscript{\ddag} & I think the AI agent brought valuable insights into potential risks \\
Q4\textsuperscript{\ddag} & Learning to work with the AI agent was easy for me \\
Q5\textsuperscript{\ddag} & I think collaborating with the AI agent felt natural \\
Q6\textsuperscript{\ddag} & I find the AI agents’ responses were clear and understandable \\
Q7\textsuperscript{\ddag} & My opinion of AI in SEE has become more positive \\
Q8\textsuperscript{\ddag} & I believe AI in SEE could benefit teams \\
Q9\textsuperscript{\ddag} & I think the AI agent can be a substitute for an expert \\
Q10\textsuperscript{\ddag} & I would be comfortable relying on AI agent’s estimations \\
Q11\textsuperscript{\ddag} & I am more likely to recommend using AI agents \\
Q12\textsuperscript{\ddag} & I would trust the estimations provided by the AI agent \\
\hline
\multicolumn{2}{|l|}{\textbf{Open-ended Questions}} \\
\hline
Q13\textsuperscript{*} & Based on your experience, what are your thoughts on the potential benefits and challenges of using AI agents in software effort estimation? \\
Q14\textsuperscript{*} & What suggestions do you have for improving the design or functionality of the AI agents to enhance their usefulness and effectiveness in estimation sessions? \\
\hline
\end{tabular}

\vspace{0.5em}
\begin{minipage}{\linewidth}
\footnotesize
\textsuperscript{\ddag} Likert scale: $\circ$ Strongly disagree, $\circ$ Somewhat disagree, $\circ$ Neutral, $\circ$ Somewhat agree, $\circ$ Strongly agree \\
\textsuperscript{*} Open-ended questions.\\
\textsuperscript{**} The term ``AI" and ``AI agent" refer to \agent.
\end{minipage}
\end{table}

\subsubsection{Results}

\begin{figure*}[ht]
    \centering
    \includegraphics[width=1\linewidth]{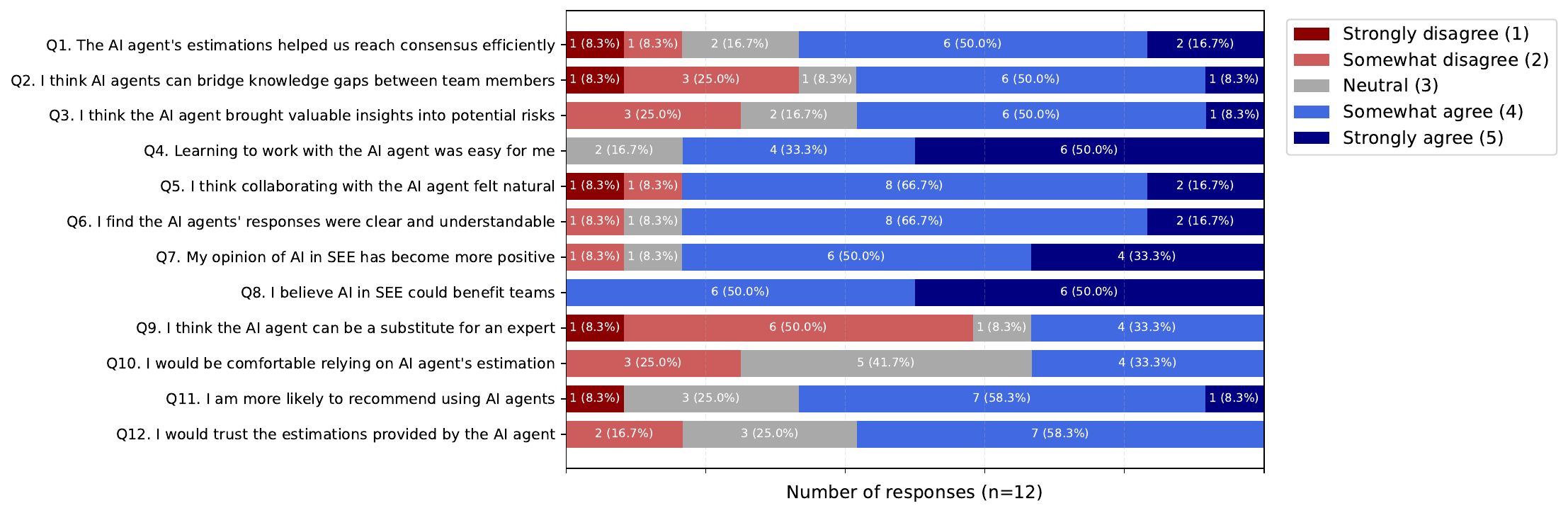}
    \caption{Survey results}
    \label{fig:quantitative-results}
\end{figure*}

Figure~\ref{fig:quantitative-results} presents the participants' responses to Questions 1--12 in our survey after they completed an agile estimation session with our \agents. 
We categorise the findings into three key dimensions: operational effectiveness, user experience, and user trust. These dimensions help understand the impact and potential of \agents in collaborating with human developers in agile effort estimation.

In terms of operational effectiveness, more than half (66.7\%) of the respondents agreed that the agents' estimations helped reach consensus more efficiently (Q1). Similarly, 58.3\% believe that the agents bridged knowledge gaps between the team members (Q2). Comparably, 58.3\% of the participants recognised the agents' capability to shed insights into potential risks (Q3). These results suggest that our \agents address some inherent challenges in agile effort estimation, particularly in terms of collective decision-making and knowledge gap management.

In addition, the results demonstrate a positive user experience with our \agents, with half of the participants finding it easy learning to work with the agents (Q4). It is also important to note that a significant proportion (83.3\%) of the participants felt natural when working with the agents' natural (Q5) and found that the agents' responses were clear and understandable (Q6).  The positive experience with our \agents has helped form a positive view towards using AI in agile effort estimation (83.3\% of agreement for Q7). These results suggest that the \agents have successfully achieved a high level of intuitiveness that could greatly facilitate their adoption in practice. The positive user experience also implies that a well-designed AI agent system can effectively mitigate professional scepticism and hesitancy toward AI adoption.

In terms of user trust, all of the participants believed that \agent would benefit agile teams in effort estimation (Q8). While only one third of them thought that the agent can fully substitute an expert (Q9), 75\% of the participants did not object to relying on the agent's estimations (Q10). The result also shows that more than half of the participants trusted \agent's estimations (Q12) and would recommend using it in practice (Q11). These results suggested that, while their utility is universally acknowledged, there remains a preference for human oversight in essential decision-making processes. This indicates that successful adoption strategies should place emphasis on humans as decision makers and AI agents as supportive collaborators.

Our analysis of the responses for the open-ended questions (Q13 and Q14) also offers further insights and confirms the positive experience of the participants in collaborating with our \agents. They found that the agents ``\textit{give us rational feedback and opinions}'' and ``\textit{benefit development teams to reach consensus}''. Many participants found the agents are useful in the agile effort estimation session since they ``\textit{bridge knowledge gap between team members (frontend vs backend dev, new vs old team members)}'' and promote communication between team members. The participants were also impressed with how our \agents used past user stories to estimate new user stories since ``\textit{it can help the team not only estimate the user stories correctly but also see the related issues arising when implementing these sorts of features in the past}''. 

The participants also suggested several improvements to \agents such as leveraging ``\emph{not only the user stories but also the source code and infrastructure}''. They also suggested integrating \agent with existing agile tools such as Jira, Trello and supporting voice communication (instead of only chat messaging) with human developers. We will develop these new features into \agents as part of our future work.

\subsection{Threats to validity}

Our selection of the foundational LLM of our \agent poses an internal threat. Different LLMs may exhibit different reasoning capabilities, which might affect the agent's behaviour in estimating user stories and collaborating with human agile team members to reach a consensus estimate. We have mitigated this threat by designing \agent's architecture such that it does not depend on any specific LLM. In addition, we have also experimented with two widely used foundational LLMs (Meta's Llama and OpenAI's GPT models). Future work would involve integrating other LLMs and investigating their impact on the performance of our agents in the agile estimation context. In addition, our replication of the SOTA baselines strictly followed the description provided in their work, however we acknowledge that the environment we set up might be different from their original environment. To mitigate this threat, we used the same dataset and maintained consistent experimental settings across all comparisons to ensure fair evaluation.

A key external threat to our human study resides in the limited sample size of 12 participants. This relatively small number of participants may not be fully representative of the broader software development community, limiting the generalisation of our study. However, we note that recruiting participants to join live collaborative estimation sessions is significantly challenging since it takes significant time and effort. This sample size of human study is also common in the software engineering literature. The result has shown the effectiveness of our \agents in collaborating with human developers in agile effort estimation. Insight from this study will help us attract the community's interest and extend it to a larger number of participants.

\section{Related work}\label{sec:related-work}

Agile effort estimation has been attracting significant interest in recent years. A range of machine learning-based techniques have been proposed to estimate user stories using the knowledge learnt from past user stories in the same project. One common approach is processing the description of a user story into a set of learning features (e.g. TF-IDF features as in \cite{porruEstimatingStoryPoints2016}). Other approaches also explore the use of developer-related features (e.g. \cite{scottUsingDevelopersFeatures2018}) to improve the estimation accuracy, or employ clustering techniques to group similar issues and use this as the basis for estimation \cite{tawosiInvestigatingEffectivenessClustering2022}. 

Deep learning approaches are also widely explored to tackle the problem since the seminal work of Choetkiertikul \emph{et al.} \cite{choetkiertikulDeepLearningModel2019}, which leverages Long Short-Term Memory (LSTM) networks and Recurrent Highway Networks (RHN) to generate semantic features from user story descriptions and use them for estimation. In 2022, Ahmad and Ibrahim used only LSTM \cite{ahmadSoftwareDevelopmentEffort2022} for SEE and yielded better accuracy over traditional approaches.
Fu and Tantithamthavorn \cite{fuGPT2SPTransformerBasedAgile2023} introduced GPT2SP, a transformer-based architecture by employs pre-trained language models and results in superior accuracy over previous methods. In 2023, Kassem \emph{et al.} \cite{kassemStoryPointEstimation2023} developed a Deep Attention Neural Network that uses the attention mechanism to capture relative word importance in issue descriptions. A recent approach called Fine-SE \cite{liFineSEIntegratingSemantic2024} explored the idea of combining expert features with semantic features to produce better estimations. While these approaches have shown promising results in terms of accuracy, they operate as black-box models that do not explain their estimation nor participate in team discussion, an important aspect of agile effort estimation. Our \agent framework leverages LLMs to create agents that not only provide estimates but also provide relevant justifications. Furthermore, they can communicate with human practitioners and provide different perspectives on a user story, similar to how cross-functional agile teams operate in practice.
    
Recent research in LLM-based agents has demonstrated their capability to autonomously solve complex and human-like tasks \cite{wuAutoGenEnablingNextGen2024,liCAMELCommunicativeAgents2023a}. Inspired by those breakthroughs, a range of approaches have been recently proposed to develop LLM-based agents for performing software engineering tasks such as requirement elicitation (e.g. \cite{ataeiElicitronLLMAgentBased2024}), code generation (e.g. \cite{liCodeTreeAgentguidedTree2024,nhatphanHyperAgentGeneralistSoftware2024}), quality assurance (e.g. \cite{dengPENTESTGPTEvaluatingHarnessing2024, yoonIntentDrivenMobileGUI2024}) and software development (e.g. \cite{taoMAGISLLMBasedMultiAgent2024, hongMetaGPTMetaProgramming2023a}).  Tao et al. \cite{taoMAGISLLMBasedMultiAgent2024} used LLM agents to simulate planning discussions for software tasks before the commencement of coding activities. Recent studies \cite{zhangEmpoweringAgileBasedGenerative2025, nguyenAgileCoderDynamicCollaborative2024} have demonstrated promising results of multi-agent systems in emulating agile software development workflows, where agents collectively simulate the roles and interactions inherent in agile methodologies. While prior research has shown the effectiveness of LLM-based multi-agent systems in various project management contexts, there remains a gap regarding their application in agile effort estimation. Our work distinguishes itself by studying not only the quantitative accuracy of estimations, but also understanding the human-AI interaction dynamics and practitioners' perceptions of AI agents in the context of a collaborative effort estimation scenario.
    
\section{Conclusions and future work}\label{sec:conclusion}

In the paper, we have proposed \agent, an LLM-based multi-agent framework which is capable of not only estimating user stories accurately but also coordinating, communicating and discussing with human developers to reach an agreed-upon estimate. Our approach represents a significant departure from the prior state-of-the-art machine learning-based techniques in agile estimation, which \emph{cannot} even explain nor justify their estimates.  
Results from a rigorous human study have shown that our \agents can engage in meaningful discussions with human practitioners, make accurate estimation on user stories and provide useful justification. The participants found that our agents help them reach consensus more efficiently, bridge the knowledge gap, and bring valuable insights into potential risks.  The participants also agreed that they would trust our \agent and recommend using it in practice.



Feedback from the participants also provide several avenues for our future work such as enriching \agent's knowledge with source code and infrastructure data, or integrating it with existing agile project management tools such as JIRA. We will also incorporate voice interaction capability into \agent to facilitate more natural communication with humans. Our future work also involves expanding the human study significantly into the commercial domain to bring our \agent to the wider agile software development communities. 



\balance
\bibliographystyle{IEEEtran}
\bibliography{main}

\end{document}